\documentclass[acus]{JAC2003}

\usepackage{graphicx}
\usepackage{booktabs}
\usepackage{amsmath}
\usepackage{multirow}

\begin{document}
\title{BEAM DYNAMICS STUDIES OF THE REX-ISOLDE LINAC IN PREPARATION FOR ITS ROLE AS INJECTOR FOR THE HIE-ISOLDE SC LINAC AT CERN}
\author{M.A. Fraser$^{\dag \ddag}$\thanks{matthew.alexander.fraser@cern.ch}, M. Pasini$^{\S \ddag}$, R.M. Jones$^{\dag}$, D. Voulot$^{\ddag}$\\
$^{\dag}$The University of Manchester, Oxford Road, Manchester, M13 9PL, UK.\\
$^{\dag}$The Cockcroft Institute, Daresbury, Warrington, Cheshire WA4 4AD, UK.\\
$^{\S}$Instituut voor Kern- en Stralingsfysica, K.U.Leuven, Celestijnenlaan 200D
B-3001 Leuven, BE.\\
$^{\ddag}$CERN, Geneva, Switzerland.\\}

\maketitle

\begin{abstract}
	The superconducting (SC) High Intensity and Energy (HIE) ISOLDE linac will replace most of the existing accelerating infrastructure of the Radioactive ion beam EXperiment (REX) at CERN, however, the 101.28 MHz RFQ and 5 MV IH cavity will remain in the role of injector for the upgrade, boosting the beam up to an energy of 1.2 MeV/u. We present the results of a beam dynamics investigation of the injector focused most critically on matching the longitudinal beam parameters from the RFQ to the SC machine, which is complicated largely by the IH cavity employing a Combined Zero Degree (KONUS) beam dynamics design,~\cite{IHS}. The longitudinal beam parameters at the RFQ are reconstructed from measurement using the three gradient method and combined with beam dynamics measurements and simulations of the IH structure to design the matching section for the SC linac.
\end{abstract}

\section{INTRODUCTION}
\begin{figure}[ht]
   \centering
   \includegraphics[width=83mm]{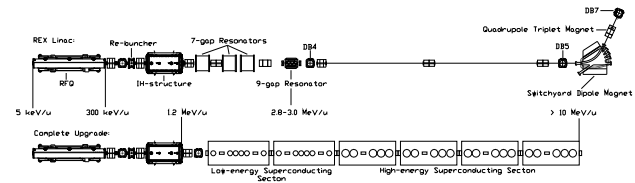}
   \caption{The layout of the REX linac (upper) and after the HIE upgrade (lower).}
   \label{layout}
\end{figure}

As was reported in~\cite{REX_EMIT}, a beam dynamics investigation of the REX linac was initiated in order to fix the input beam parameters for the design of the HIE upgrade. The longitudinal emittance was measured by means of the three gradient method behind the RFQ at 300 keV/u and the second 7-gap resonator at 1.92 MeV/u using the re-buncher and the third 7-gap resonator in a bunching mode and by combining the switchyard dipole magnet and quadrupole on the 65$^\circ$ beam line to form a spectrometer, as is presented in Fig.~\ref{layout}. The upgrade will proceed first with the addition of high energy SC cyromodules after the 9-gap resonator before the installation of a low energy section, also shown in the Fig.~\ref{layout}.

\section{LONGITUDINAL EMITTANCE MEASUREMENTS}
Using residual gas from the electron beam ion source, composed predominantly of $\textrm{Ne}^{5+}$ leaked from the adjacent Penning trap, at $A/q = 4$ some tens of pA of beam current was accelerated to either the re-buncher or the third 7-gap resonator and then transported to the spectrometer. At 300 keV/u only 50~\% of the beam entering the RFQ could be transmitted along the linac whereas at 1.92 MeV/u over 80~\% was achieved. Using a \texttt{TRACE3D} model an astigmatic spectrometer system was set up using the first quadrupole behind the dipole to image in the dispersive plane the vertical slit in diagnostic box (DB) 5 through the switchyard magnet and onto the Faraday cup in DB7. A 1~mm vertical slit in DB5 gives a calculated energy resolution of 0.18~\% on the 65$^\circ$ beam line but limits the intensity on the Faraday cup in DB7 to just a few pA. In order to profile the beam at such low intensities an aperture larger than the horizontal beam size was used in front of the Faraday cup and the beam scanned across it by varying the magnetic field of the bender. The current on the Faraday cup was acquired as a function of the dipole field, calibrated at the RFQ output energy of 300 keV/u, and the beam profile reconstructed on each side of the aperture from the rate of change of the current as the beam moved across the edges of the aperture. Profiling the beam in this way perturbs the optics of the system. However, a variation of just $\pm2$~\% in the dipole field was needed to scan the beam across a 15 mm aperture and as a result, at 300 keV/u, the resolution is calculated to increase from 0.18 to 0.32~\%, assuming a beam divergence of 10~mrad at the slit in DB5. 
\begin{figure}[htb]
   \centering
   \includegraphics*[width=70mm]{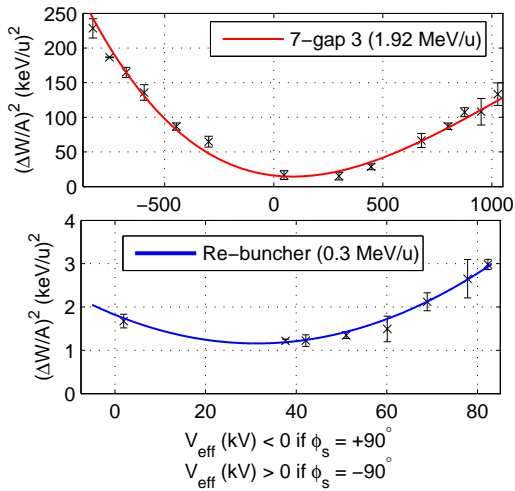}
   \caption{The rms energy spread after the re-buncher (lower) and third 7-gap resonator (upper), fitted with quadratic and quartic functions of effective voltage for emittance measurement.}
   \label{rms_data}
\end{figure}The re-buncher is well approximated by a thin bunching cavity and the longitudinal emittance can be measured using the standard three gradient technique employing a quadratic fit of the square of the energy spread as a function of the effective voltage of the buncher,~\cite{REX_EMIT,SLAC_EMIT}. This technique was extended in order to measure the emittance after the IH structure using the third 7-gap resonator, which is a split-ring structure with a geometric velocity of 6.6~\%. The longitudinal transfer matrix element describing the evolution of the energy spread in the 7-gap resonator was derived assuming that the beam velocity is matched to the structure so that the beam experiences the same rf phase in each gap. If one keeps only linear terms of the ratio of the effective voltage to beam energy ($\frac{V_{eff}}{W_0}$) in the transfer matrix one arrives at a quartic equation in $V_{eff}$ that describes the energy spread after the multi-gap structure as a function of the input Twiss parameters,~\cite{rex_emit_note}. The data are fitted in Fig.~\ref{rms_data}, the beam parameters collected in Tab.~\ref{paras} and the measured rms beam ellipse behind the RFQ compared to \texttt{PARMTEQ} simulations in Fig.~\ref{beam}. Simulation shows that the longitudinal emittance can be reconstructed accurately at low voltages and to within 15~\% for $V_{eff} < 1.1$~MV using the 7-gap cavity and to the same accuracy for  $V_{eff} < 80$~kV using the re-buncher. The ratio $\frac{\alpha}{\beta}$ is independent of resolution, however, the number of data points that could be taken around the minimum of the energy spread was limited by the achievable stable field level in the re-buncher at low power and as a result the fit produces a large error in this parameter. It is foreseen to repeat the measurement with a higher $A/q$ to improve the uncertainty.\begin{table}[hbt]
   \centering
   \footnotesize
   \caption{Longitudinal beam parameters in front of the 7-gap 3 and the re-buncher, where the measurement at 300 keV/u is compared to simulation.}
   \begin{tabular}{l|c|ccc}
       \toprule
         \textbf{Data Set}   & \textbf{7-gap 3} & \textbf{Re-buncher}& \textbf{PARMTEQ} \\ 
         &  &  & \textbf{Sim.} \\ 
       \midrule
         $\alpha$       &$0.14\pm0.07$  &$0.76\pm0.15$  & 2.3  \\
         $\beta$ (ns/keV/u)        &$0.022\pm0.004$       & $0.15\pm0.02$ & 0.6   \\
         $\frac{\alpha}{\beta}$ (keV/u/ns)    &    $6.4\pm3.4$     &  $5.0\pm1.1$ &  3.8 \\
          $\epsilon_{rms}$ ($\pi$ ns keV/u)& $0.35\pm 0.04$&$0.18\pm 0.02$ & 0.26   \\
         \bottomrule
   \end{tabular}
   \label{paras}
\end{table}
\begin{figure}[htb]
   \centering
   \includegraphics*[width=60mm]{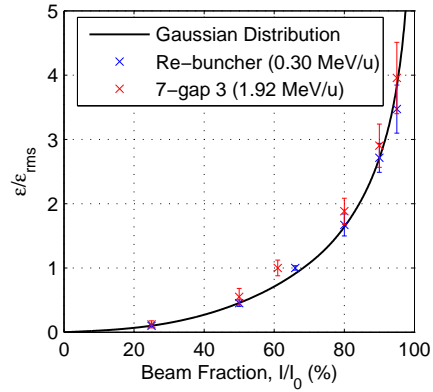}
   \caption{The measured longitudinal distribution of the beam compared to a Gaussian distribution.}
   \label{dist}
\end{figure}The measured rms and 95~\% emittances at RFQ energy are respectively 30~\% and 60~\% lower than predicted, which, when compared to the measurement using the third 7-gap cavity, suggests that the loss in transmission between the re-buncher and the spectrometer was correlated to energy. The transmission through the spectrometer was better than 95~\%. At 1.92 MeV/u the measured rms emittance is larger than predicted, which can be attributed to the measurement procedure itself and the resolution of the spectrometer. In any case, the effect of the resolution increases the measured value of the rms emittance and the values presented here can therefore be taken as upper limits. As shown in Fig.~\ref{dist}, the longitudinal distribution of the beam is closely Gaussian and the effect of the IH and 7-gap cavities is to systematically redistribute particles away from the beam core: the rms emittance at 0.3~MeV/u contains 66~\% of the beam whereas at 1.92~MeV/u it contains just 61~\%. The 95~\% emittance measured using the 7-gap cavity is 1.56~$\pi$~keV/u~ns, which is smaller than the simulated value of 1.76~$\pi$~keV/u~ns and ascribed to the losses. The total design value of the emittance used in the beam dynamics simulations of the SC linac is 2~$\pi$~keV/u~ns.

\section{SIMULATIONS AND MEASUREMENTS OF THE IH STRUCTURE}
\begin{figure}[htb]
   \centering
   \includegraphics*[width=60mm]{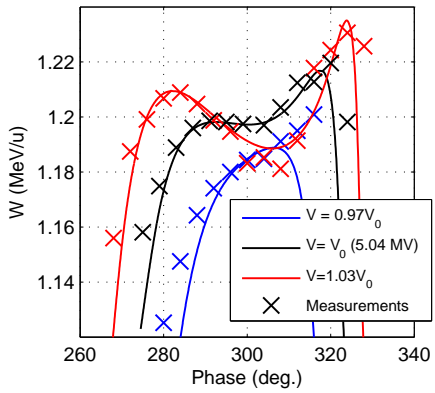}
   \caption{Phase-up measurements of the IH structure contrasted to simulation at different field amplitudes.}
   \label{comparison}
\end{figure}

After successful bead pull measurements of the IH structure the encoder values of the plunger tuners were re-calibrated and referenced with respect to external fiducials for an accelerating profile delivering beam at 1.2~MeV/u. The measured accelerating field was shown to be in excellent agreement with an electromagnetic simulation of the structure using \texttt{HFSS},~\cite{REX_EMIT}. The simulated field map was applied to benchmark the \texttt{LORASR} code used to design the IH cavity and to track the realistic particle distribution from the RFQ exit to the entrance of the SC linac. In order to ensure the correct amplitude and phase settings without an extensive recommissioning the measured average energy gain of the beam as a function of rf phase was compared to simulation, as shown in Fig.~\ref{comparison}. The IH structure was under and over powered by 9~\% and the response was well predicted by simulation. At nominal field level the phase is set to 299$^\circ$. 
The simulation also allows the systematic error on the calibration of the spectrometer at RFQ energy to be estimated at 1.4~\% from the offset in energy required to match the measurement to the simulation. The shape of the peak in Fig.~\ref{comparison} is highly sensitive to the injection energy and field amplitude in the cavity and is therefore a good indicator of whether the cavity settings are close to the design values.

\section{DESIGN OF THE MATCHING SECTION}
\begin{figure}[htb]
   \centering
   \includegraphics*[width=82mm]{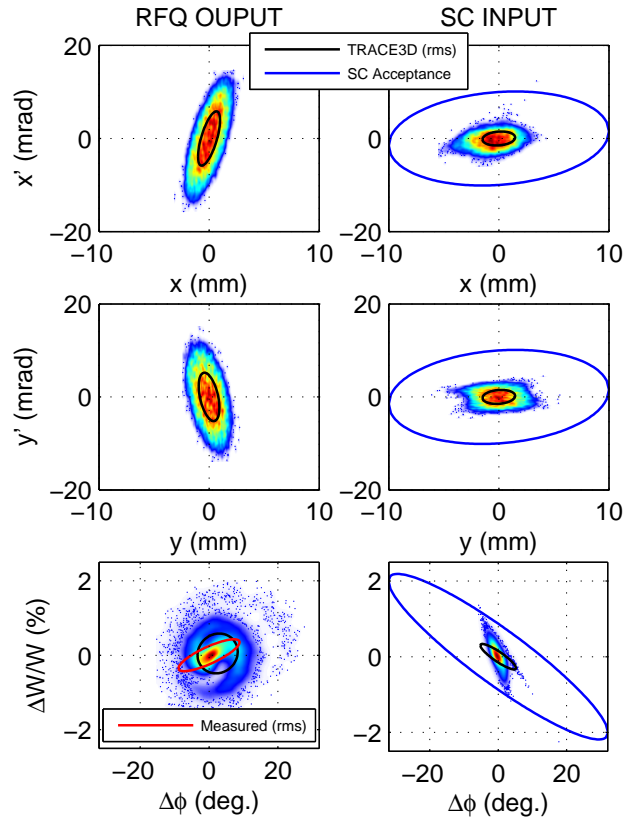}
   \caption{The beam tracked from behind the RFQ through the REX injector to the first cavity of the SC linac.}
   \label{beam}
\end{figure}

\begin{figure}[htb]
   \centering
   \includegraphics*[width=70mm]{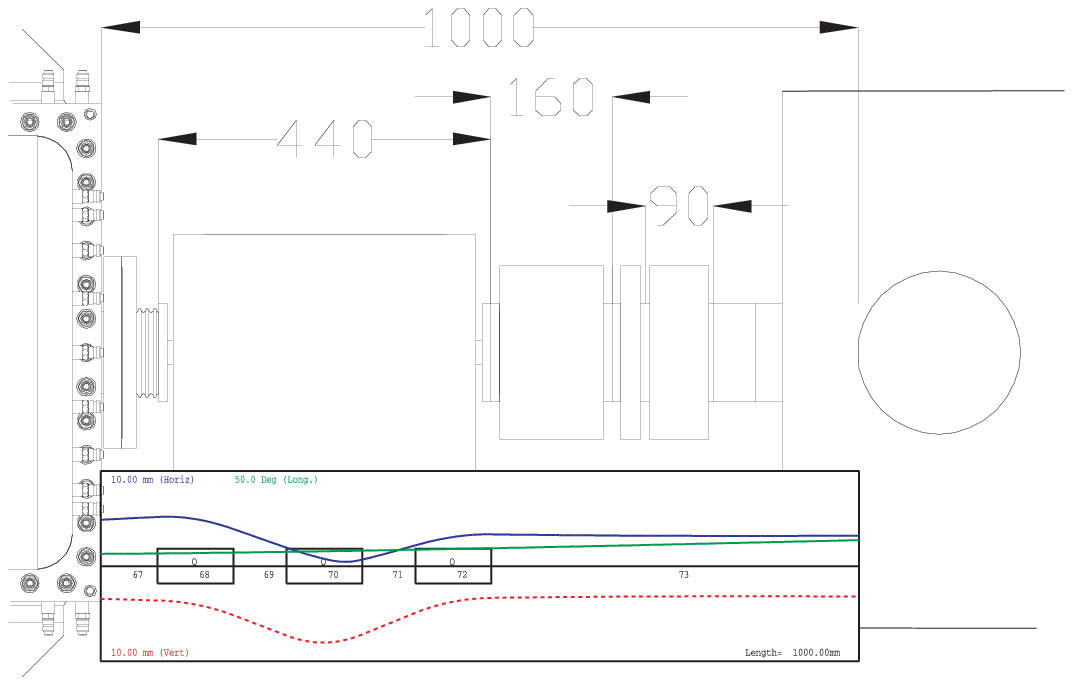}
   \caption{Matching section components left to right: IH, quadrupole triplet, warm steerer, cold trap, diagnostic box, valve, SC cavity. Dimensions in mm.}
   \label{m_section}
\end{figure}

The matching section was designed using a \texttt{TRACE3D} model of the REX linac benchmarked against simulations which tracked the realistic particle distribution from the output of the RFQ, through the field maps of the re-buncher and IH structure, to the SC linac as shown in Fig.~\ref{beam}. The beam is divergent in both transverse planes at output from the IH structure and can be matched symmetrically into the SC solenoid channel within a distance of 1~m using a quadrupole triplet magnet. The compact nature of the matching section ensures that the beam can be re-captured longitudinally by the first cavity of the low energy SC section operated in a bunching mode. The third quadrupole of the triplet inside of the IH cavity can be used for fine tuning. A pole tip radius of 15 mm is adequate to keep 95~\% of the beam within half the aperture demanding up to 0.9~T on the pole tip. The design is summarised in Fig.~\ref{m_section}.
\section{CONCLUSION}
A novel measurement of the longitudinal emittance of the REX-ISOLDE linac was presented and the results combined with measurements and simulations of the IH cavity to design the matching section for the SC HIE-ISOLDE upgrade.
\section{ACKNOWLEDGEMENTS}
The authors would like to thank Piero Posocco and Emiliano Piselli for their support throughout the campaign of measurements and simulations.

\end{document}